\documentclass[prd,aps,reprint,groupedaddress,floatfix,flushbottom,nobibnotes,nofootinbib,showpacs,amssymb,amsmath]{revtex4-1}
\usepackage{graphicx}
\usepackage{longtable}

\begin{document}
\title{The moment 
$\langle x\rangle_{u-d}$
of the nucleon from $N_f=2$ lattice QCD down to nearly physical quark masses}
\author{Gunnar S.~Bali}
\author{Sara~Collins}
\email{sara.collins@ur.de}
\author{Benjamin~Gl\"a\ss{}le}        
\author{Meinulf~G\"{o}ckeler}        
\author{Johannes Najjar}        
\author{Rudolf~H.~R\"odl}  
\author{Andreas~Sch\"{a}fer}  
\author{Rainer~W.~Schiel}  
\author{Andr\'e~Sternbeck}  
\author{Wolfgang~S\"oldner}  
\affiliation{Institut f\"ur Theoretische Physik, Universit\"at Regensburg,
              93040 Regensburg, Germany}   
\date{\today}
\begin{abstract}
We present an update of our analysis \cite{Bali:2012av} which includes
additional ensembles at different quark masses, lattice spacings and
volumes, all with high statistics.  We use $N_f=2$ mass-degenerate quark flavours,
employing the non-perturbatively improved clover action.  The lattice
matrix elements are converted to the $\overline{\rm MS}$ scheme via
renormalization factors determined non-perturbatively in the
RI$^\prime$-MOM scheme. We have systematically investigated excited
state contributions, in particular, at the smallest, near physical,
pion mass. While our results~(with much increased precision) are
consistent with Ref.~\cite{Bali:2012av}, comparing with previous
determinations we find that excited state contributions can be
significant if the quark smearing is not suitably optimized, in
agreement with other recent studies.
The difference with respect to the value for $\langle x\rangle_{u-d}$
extracted from experimental data is reduced but not resolved. Using
lattice sizes in the range $L m_\pi\sim 3.4-6.7$, no significant finite
volume effects were observed.
Performing a controlled continuum limit that may remove the discrepancy
will require simulations at lattice spacings $a< 0.06$~fm.
\end{abstract}
\maketitle

\section{Introduction}
The focus of modern hadron structure physics has evolved from ordinary
parton distribution functions (PDFs) and form factors to more complex
quantities like generalized parton distributions (GPDs), transverse
momentum dependent PDFs (TMDs), double distributions (DDs) and
distribution amplitudes (DAs), see e.g.\ Ref.~\cite{Boer:2011fh}.
Many of these quantities cannot easily be accessed by experiment and
lattice results often have to be used as a substitute.  In this
situation, observables where lattice results and experiment can be
compared to judge the reliability of lattice predictions~(and of
phenomenological fits to experimental data) play a key role.  In
recent years good agreement has been found for such benchmark
quantities, e.g., in hadron
spectroscopy~\cite{Durr:2008zz,Fodor:2012gf}, however, for a
fundamental and well-known nucleon structure observable, namely the
iso-vector quark momentum fraction, $\langle x\rangle_{u-d}$, significant disagreement
remains.  This needs to be resolved to improve prospects for hadron
physics beyond the level of PDFs. For recent reviews concerning
lattice hadron structure determinations see
e.g. Refs.~\cite{Syritsyn:2014saa,Brambilla:2014aaa,Alexandrou:2013asa}.

Past lattice predictions were complicated by the
need to extrapolate lattice results determined at larger than
physical quark masses using, e.g., parameterizations given by
chiral perturbation theory~(ChPT).  The range of validity of ChPT
depends on the quantity studied and can ultimately only be tested by
lattice simulations, including the physical point. In addition, any
extrapolation is unreliable if the lattice results themselves do not
include reasonable estimates of their main systematics.  Consequently,
we started a dedicated effort to produce high statistics results for
$N_f=2$ fermions at nearly physical quark masses. 

This began with a study at $m_\pi\lesssim 160$~MeV on a single small
volume, with linear lattice extent $L$ in units of the pion mass
$L m_\pi\sim 2.77$, detailed in Ref.~\cite{Bali:2012av}. With the
aim of investigating the main sources of systematic uncertainty we
have since expanded our data set to include a larger volume with
$m_{\pi}\sim 150$~MeV and $L m_\pi\sim 3.49$ and several ensembles at
larger quark masses~(up to $m_\pi\sim 490$~MeV), a range of volumes~($L m_\pi
\sim 3.4-6.7$) and a limited range of lattice
spacings~($a\sim 0.06-0.08$~fm). In addition, we have performed a thorough
analysis of excited state contributions, which $\langle
x\rangle_{u-d}$ is known to be sensitive
to~\cite{Bali:2012av,Dinter:2011sg,Owen:2012ts,Capitani:2012gj,Green:2012ud,Bhattacharya:2013ehc}.
Nevertheless we find our results for $\langle x\rangle_{u-d}$
still to deviate from the phenomenological values by roughly 25\%.

The structure of this paper is as follows: After detailing our lattice
set up in Section~\ref{latsetup} we discuss our analysis of excited
states in Section~\ref{excitedstates} and present consistency checks
performed involving finite momentum data in
Section~\ref{finitemom}. Our final results are compared with other
recent determinations in Section~\ref{sys} and we discuss remaining
systematics in the conclusions, Section~\ref{conc}. We note that a preliminary
analysis of some of our ensembles appeared in
Ref.~\cite{Bali:2013nla}.

\section{Lattice set up}
\label{latsetup}

We used configurations generated by the Regensburg QCD collaboration~(RQCD) and QCDSF
with $N_f=2$ non-perturbatively improved clover fermions and the
Wilson gauge action, see Table~\ref{tab_1} for the simulation
parameters.  While many lattice simulations now also include dynamical
strange quarks, so far strangeness has been found to play a minor role
in nucleon structure~\cite{Doi:2009sq,QCDSF:2011aa,Bali:2011ks,Abdel-Rehim:2013wlz} and $N_f = 2$ simulations
remain relevant.

\begin{table*}
\caption{Overview of ensembles used for this analysis. $N(n)$
  indicates the number of configurations, $N$, and the number of
  measurements, $n$, per configuration of the two- and three-point
  functions. Statistical noise decreases with decreasing $t_{\rm f}$
  and for some of the three-point functions we used a smaller number
  of measurements per configuration as indicated in brackets in the
  next-to-last column of the table. $N_{sm}$ refers to the number of
  iterations used for Wuppertal smearing.  Note that the errors of
  the~(finite volume) pion masses combine the statistical uncertainty
  with an estimate of the variation in the mass arising from the
  choice of fitting range. For $\langle x\rangle_{u-d}^{\overline{\rm
      MS}}(2\,\mathrm{GeV})$, the error includes the same sources of
  uncertainty and in addition the error associated with the
  renormalization factors.  }\label{tab_1}
\begin{ruledtabular}
\begin{tabular}{c|ccccccccccc}
ensemble& $\beta$ &  $a$ [fm] & $\kappa$     &   $V$   & $am_\pi$ & $m_\pi$ [GeV] &  $L m_\pi$  &   $N(n)$&  $N_{sm}$ & $t_{\rm f}/a$ & $\langle x\rangle_{u-d}^{\overline{\rm MS}}(2\,\mathrm{GeV})$ \\
\hline
I&5.20 &  0.081& 0.13596 &  $32^3\times 64$ &  0.11516(73) & 0.2795(18)	&    3.69 &  $1986(4)$ &   300 &    13  & 0.195(10) \\
\hline
II&5.29 &  0.071 & 
0.13620 &   $24^3\times 48$  & 0.15449(74)    & 0.4264(20)  &     3.71   &   $1999(2)$   &  300  &   15  & 0.230(10)  \\
III&&&0.13620  &   $32^3\times 64$  &  0.15298(46) &   0.4222(13)   &      4.89   &   $1998(2)$   &  300  &   15,17 & 0.215(04) \\
IV&&&0.13632 &    $32^3\times 64$ & 0.10675(51)  & 0.2946(14) &      3.42   &   $2023(2)$   &  400   &  7(1),9(1),11(1), & 0.206(09) \\
&&&&    &          &         &             &            &       & 13,15,17 & \\
V&&&&   $40^3\times 64$  &  0.10465(38) &  0.2888(11)    &      4.19   &   $2025(2)$   &  400  &   15 & 0.218(08) \\
VI&&&&   $64^3\times 64$  &  0.10487(24)  & 0.2895(07)    &      6.70   &   $1232(2)$   &  400  &   15 & 0.196(06)\\
VII&&&0.13640 &    $48^3\times 64$ &  0.05786(55)   & 0.1597(15)  &      2.77   &   $3442(2)$   &  400   &  15 & 0.200(12) \\
VIII&&&&   $64^3\times 64$ &  0.05425(49) & 0.1497(13)   &       3.49   &   $1593(3)$   &  400   &  9(1), 12(2), 15  & 0.217(09) \\
\hline
IX&5.40 & 0.060 &  
0.13640  &   $32^3\times 64$  &  0.15020(53) &  0.4897(17)     &   4.81   &   $1123(2)$   &  400  &   17  & 0.216(07)\\
X&&&0.13647  &   $32^3\times 64$  &  0.13073(62) & 0.4262(20)	& 4.18         &   $1999(2)$&     450 &    17  & 0.212(06)\\       
XI&&&0.13660  &   $48^3\times 64$  &  0.07959(27)  & 0.2595(09)  &   3.82   &   $2177(2)$   &  600  &   17  & 0.196(08)
\end{tabular}
\end{ruledtabular}
\end{table*}

The two-point and three-point functions, needed to extract the
quark momentum fraction have the form
\begin{eqnarray}
C_{{\rm 2pt}}(t_{\mathrm{f}}) &= & \sum_{\vec{x}}\langle  {\cal N}(\vec{x},t_{\mathrm{f}})\overline{ {\cal N}}(\vec{0},0)\rangle \\
C_{{\rm 3pt}}(t_{\mathrm{f}},t) & = &\sum_{\vec{x},\vec{y}}\langle {\cal N}(\vec{x},t_{\mathrm{f}}) O(\vec{y},t)\overline{ {\cal N}}(\vec{0},0)\rangle
\end{eqnarray}
for a nucleon, ${\cal N}$, created at a time $t_{\mathrm i}=0$, destroyed at a time $t_{\mathrm{f}}$
and with an operator $O$ inserted at a time  $t$.
In the limit of large Euclidean time separations, $t_{\mathrm{f}}\gg t\gg 0$, one obtains
\begin{eqnarray}
C_{{\rm 2pt}}(t_{\mathrm{f}}) &= &  |Z_0|^2 e^{-m_0 t_{\mathrm{f}}} + |Z_1|^2e^{- m_1 t_{\mathrm{f}}}+\ldots\label{twofit}\\
C_{{\rm 3pt}}(t_{\mathrm{f}},t) & = &
 |Z_0|^2\langle  N_0|O|N_0\rangle e^{-m_0 t_{\mathrm{f}}}\nonumber\\
&&+ Z_1^*Z_0 \langle N_1|O|N_0\rangle e^{-m_0 t}e^{-m_1(t_{\mathrm{f}}-t)}  \nonumber\\
& & + Z_0^*Z_1 \langle N_0|O|N_1\rangle e^{-m_1 t} e^{-m_0(t_{\mathrm{f}}-t)}
\nonumber\\&& + |Z_1|^2\langle N_1|O|N_1\rangle e^{-m_1 t_{\mathrm{f}}}+\ldots,
\label{threepoint}
\end{eqnarray}
where $Z_i=\langle N_i|{\cal \overline{N}}|0\rangle$ are the overlaps of the
state ${\cal \overline{N}}|0\rangle$, created by a nucleon interpolator
${\cal \overline{N}}$ with the ground and
first excited states $|N_0\rangle$ and $|N_1\rangle$,
respectively. We denote the corresponding masses as $m_0$ and $m_1$.  The
``$\ldots$'' indicate the neglected higher excitations.  The target matrix
element 
\begin{equation}\langle N_0|O|N_0\rangle = -\langle x\rangle^{\rm LAT}_{u-d}
m_0, \label{div_mass}
\end{equation}
where~($y=(\vec{y},t)$)
\begin{equation}
O=\sum_{\vec{y}}
\bar{u}_y\left(\gamma_4\!\stackrel{\leftrightarrow}{D}_4 - \frac13\boldsymbol{\gamma}
\cdot\!\stackrel{\leftrightarrow}{\mathbf{D}}\right)u_y - 
\bar{d}_y\left(\gamma_4\!\stackrel{\leftrightarrow}{D}_4 -\frac13\boldsymbol{\gamma}
\cdot\!\stackrel{\leftrightarrow}{\mathbf{D}}\right)d_y,
\end{equation}
can be extracted more easily if the ground state contribution
dominates $C_{{\rm 3pt}}$~(and similarly $C_{{\rm 2pt}}$). The
covariant derivative is defined as
$\stackrel{\leftrightarrow}{D}_\mu=\frac{1}{2}\left(\stackrel{\rightarrow}{D}_\mu-\stackrel{\leftarrow}{D}_\mu\right)$. For
each ensemble the quark smearing was optimized to minimize the excited
state contributions to the nucleon two-point function. We used gauge
invariant Wuppertal~\cite{Gusken:1989ad,Gusken:1989qx} smeared quark
sources and sinks with APE smoothed spatial gauge links
$\overline{U}_{x,j}$~\cite{Falcioni:1984ei}. The Wuppertal algorithm
involves $N_{sm}$ iterations of
\begin{eqnarray}
q^{(n)}_x &=& \frac{1}{1+6\delta}\left[q^{(n-1)}_x \right.\nonumber\\
&&\hspace{.5cm}+\left.\delta \sum_{j=1}^3
\left(\overline{U}_{x,j}q^{(n-1)}_{x+\hat{j}}+ \overline{U}^\dagger_{x-\hat{j},j} q^{(n-1)}_{x-\hat{j}}\right) \right],
\end{eqnarray}
where $n$ labels the iteration number and we choose $\delta = 0.25$.
$N_{sm}$ for each ensemble is given in Table~\ref{tab_1}. Naively, at
equal pion mass, $N_{sm}\propto 1/a^2$.  In addition, for ensemble IX,
we also computed the two and three-point functions using Jacobi
smearing~\cite{Allton:1993wc}, for different numbers of iterations,
with and without APE smoothed links, for comparison.  The Jacobi
algorithm is given by
\begin{equation}
q^{(n)}_x = q_x^{(0)}+\kappa \sum_{j=1}^3
\left(U_{x,j}q^{(n-1)}_{x+\hat{j}}+ U^\dagger_{x-\hat{j},j} q^{(n-1)}_{x-\hat{j}}\right) 
\end{equation}
where we used $\kappa=0.21$.

The three-point functions were generated using the standard sequential
propagator method~\cite{Maiani:1987by} which involves fixing
$t_{\mathrm{f}}$. Alternative approaches using stochastic estimates
have also been investigated recently, see
Refs.~\cite{Bali:2013gxx,Alexandrou:2013xon}. The
value of $t_{\mathrm{f}}$ was optimized using ensemble IV. From the last
term in Eq.~(\ref{threepoint}), one can see that $C_{{\rm 3pt}}$ may
contain contributions proportional to $\langle N_j|O|N_j\rangle$, $j\ge 1$, which
can only be resolved by varying $t_{\mathrm{f}}$. As discussed below,
by using $t_{\mathrm{f}}/a$ in the range $7-17$ we found that for our
choice of smearing the last term in Eq.~(\ref{threepoint}) is
sufficiently suppressed at $t_{\mathrm{f}}\approx 1.1$~fm~$\approx 15a $
for $\beta=5.29$ within given statistics.  This value was then
used for all $\beta=5.29$ ensembles and rescaled to
$t_{\mathrm{f}}=13a$ and $17a$ for $\beta=5.20$ and $\beta=5.40$,
respectively. As an additional check at the lightest mass point we
generated $C_{{\rm 3pt}}$ with $t_{\rm f}/a=9,12$ and $15$.  Multiple
measurements were performed on each configuration for all ensembles
and autocorrelations were investigated by binning the data.

The lattice matrix elements are converted to the $\overline{{\rm MS}}$
scheme at a scale $\mu=2$~GeV using renormalization factors determined
non-perturbatively in the RI$^\prime$-MOM scheme
in Refs.~\cite{Gockeler:2010yr,Constantinou:2013ada}~(and 3-loop continuum
perturbative factors relating the RI$^\prime$-MOM and $\overline{{\rm
    MS}}$ schemes).  Applying $O(a)$ Symanzik improvement~(as was
implemented for the quark action)
the relation between
the renormalized and lattice operators has the form~\cite{Capitani:2000xi}
\begin{eqnarray}
& O^{\overline{{\rm MS}}}(\mu) & = Z^{\overline{{\rm MS}},{\rm LAT}}(a\mu)\left[(1+b am_q)O^{\rm LAT} + ac_O O_1^{\rm LAT} \right]\nonumber\\
&&\label{renorm_eq}
\end{eqnarray}
with $am_q=\frac{1}{2}\left(\kappa^{-1}-\kappa_{c}^{-1}\right)$,
  where $\kappa_c=0.1360546(39)$, $0.1364281(12)$ and
  $0.13667928(108)$ for $\beta=5.20$, $5.29$ and $5.40$, respectively.
  We set $b=1$ and $c_O=0$, thus, our values for $\langle x\rangle_{u-d}$ still
  have $O(a)$ leading discretization errors.

\section{Suppression of excited states} 
\label{excitedstates}

\begin{figure}[t]
\centerline{
\includegraphics[width=.48\textwidth,clip=]{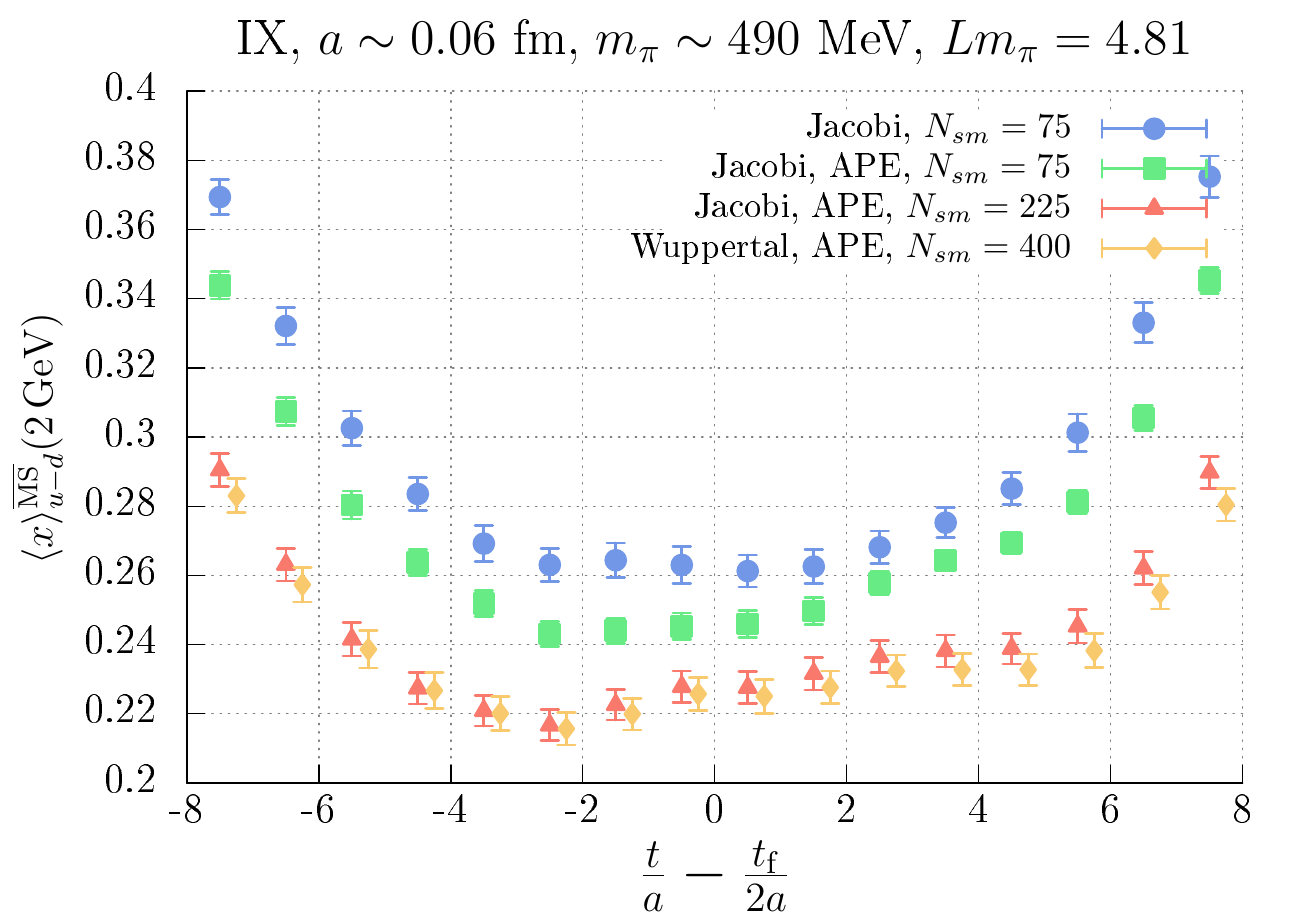}
}
\caption{Results for the ratio of three- and two-point functions for
  ensemble IX using different smearing algorithms, with and without
  APE smoothed links and for different numbers of smearing
  iterations. The ratio is multiplied by the appropriate
  renormalization factor to give $\langle
  x\rangle_{u-d}^{\overline{\rm MS}}(2\,\mathrm{GeV})$ in the limit of
  ground state dominance. The Wuppertal smeared results have been
  shifted horizontally for clarity. }
\label{fig_A21}
\end{figure}

In the past few years a number of studies have highlighted excited
state contamination as one of the main systematic uncertainties in
lattice determinations of $\langle
x\rangle_{u-d}$~\cite{Bali:2012av,Dinter:2011sg,Owen:2012ts,Capitani:2012gj,Green:2012ud,Bhattacharya:2013ehc}. Fig.~\ref{fig_A21},
which shows the ratio 
\begin{equation}
C_{{\rm 3pt}}(t,t_{\rm f})/C_{{\rm 2pt}}(t_{\rm f}) = \langle N_0|O|N_0\rangle + \ldots
\label{ratiothreetwo}
\end{equation}
for different smearings, illustrates the difficulty in extracting the
contribution of the ground state matrix element. The ratio is
multiplied by the renormalization factors in Eq.~(\ref{renorm_eq}) and
divided by the ground state mass~(cf. Eq.~(\ref{div_mass})) to give $\langle
x\rangle_{u-d}^{\overline{\rm MS}}(2\,\mathrm{GeV})$ in the region of
ground state dominance.
This ratio is
symmetric about $t=t_{\rm f}/2$ and one may mis-identify ground
state dominance, e.g., for the results with Jacobi smearing and
$N_{sm}=75$ if no other data are available. One can see that using APE
smoothed links improves the overlap with the ground state and, as
might be expected, both Wuppertal and Jacobi smearing give
compatible results once the number of smearing iterations is high
enough.

If the ground state is not dominant, similar problems arise if only
one value of $t_{\rm f}$ is available, in particular, since, as
mentioned previously, there are terms that cannot be resolved without
varying $t_{\rm f}$.  With the sequential source method the
computational expense increases linearly with the number of
$t_{\rm f}$ values~(and similarly if the smearing is varied). However,
with the computing resources now generally available, such studies have become
possible.

\begin{figure}[t]
\centerline{
\includegraphics[width=.48\textwidth,clip=]{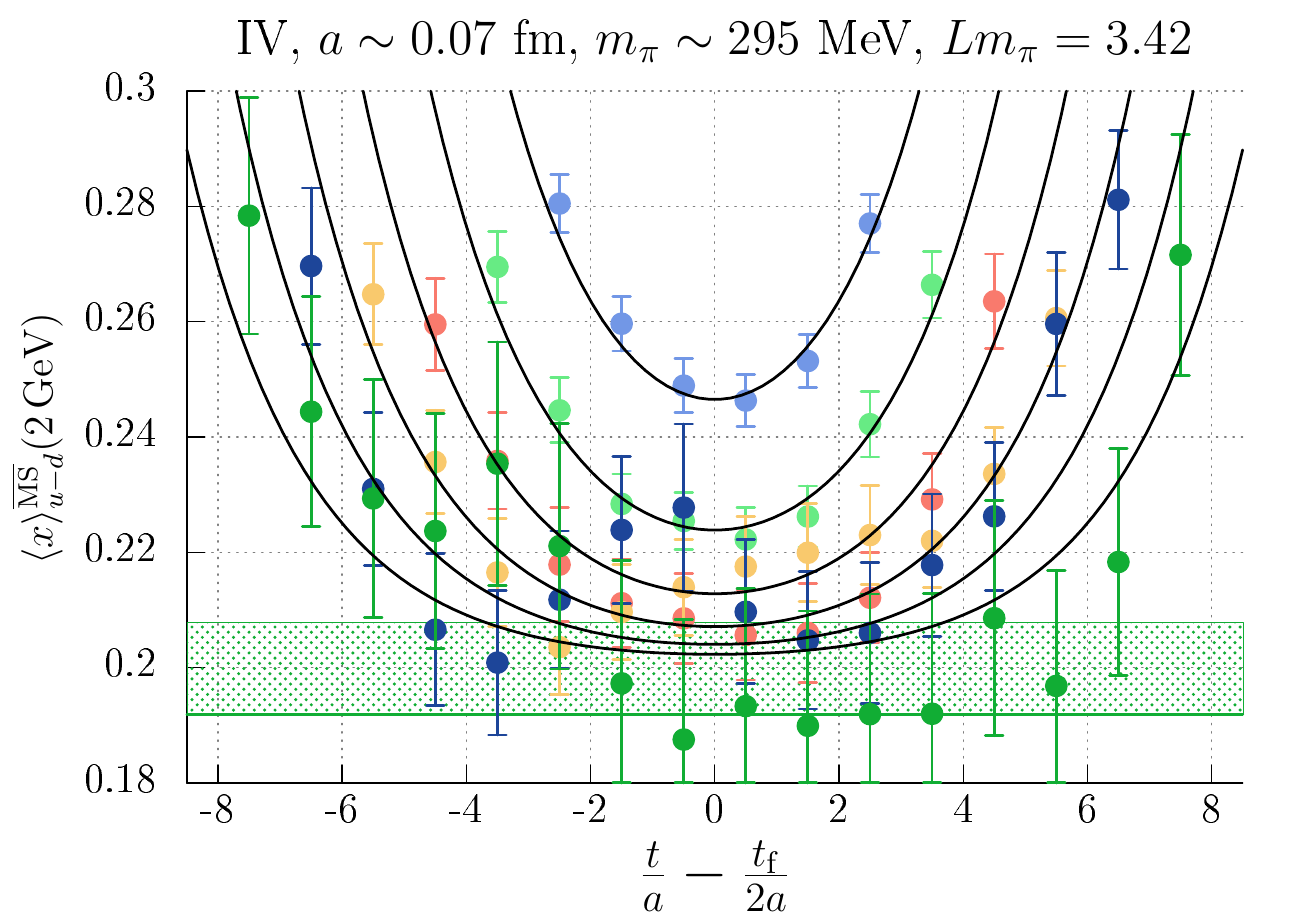}}
\caption{For ensemble IV~($m_\pi\sim 295$~MeV), the combination
  $C_{{\rm 3pt}}(t,t_{\rm f})/(m_0A_0e^{-m_0t_{\rm f}})$  times the renormalization factor
  to give $\langle
  x\rangle_{u-d}^{\overline{\rm MS}}(2\,\mathrm{GeV})$ 
for $t_{\rm
    f}/a= 7,9,11,13,15,17$.
$A_0e^{-m_0t_{\rm f}}$ corresponds to the
  ground state contribution to $C_{{\rm 2pt}}(t_{\rm f})$ obtained
  from a simultaneous fit to $C_{{\rm 3pt}}$ and $C_{{\rm 2pt}}$ for
  all $t_{\rm f}$.  Also shown is the combined fit for
  each $t_{\rm f}$. The green shaded region indicates the fitted value
  of $\langle x\rangle_{u-d}^{\overline{\rm MS}}(2\,\mathrm{GeV})$ and the
  corresponding statistical uncertainty, from a fit in the range
  $t_{\rm min}-t_{\rm max}=2a-26a$ for $C_{\rm 2pt}$ and $\delta t=2a$ for
  $C_{\rm 3pt}$.  }
\label{fig_A4}
\end{figure}

Our strategy was to minimize excited state contributions by optimizing
the smearing for the nucleon source/sink operators and to investigate
residual contamination using a range of $t_{\rm f}$ values. This
analysis was performed for ensemble IV~($m_\pi\sim 295$~MeV) with
$t_{\rm f}/a=7,9,11,13,15,17$, and 
ensemble VIII~($m_\pi\sim 150$~MeV) with $t_{\rm f}/a=9,12,15$, where $15a\sim 1.1$~fm. 
We perform simultaneous fits to the two- and three-point correlators
for all $t_{\rm f}$s using a functional form which includes the first
excited state:
\begin{eqnarray}
C_{{\rm 2pt}}(t_{\rm f}) &=& A_0e^{-m_0t_{\rm f}}+A_1e^{-(m_0+\Delta m) t_{\rm f}}, \label{twopta}\\
C_{{\rm 3pt}}(t,t_{\rm f}) &=& A_0e^{-m_0t_{\rm f}} \Bigl[ B_0+ B_1 \Bigl(e^{-\Delta m(t_{\rm f}-t)}  +  e^{-\Delta m \,t}\Bigr) \nonumber\\
& & 
\hspace{1.7cm} + B_2 e^{-\Delta m\, t_{\rm f}}\Bigr],\label{threepta}
\end{eqnarray}
where $\Delta m = m_1-m_0$, $B_0= \langle N_0|O|N_0\rangle$ and 
\begin{equation}
B_1 = \sqrt{\frac{A_1}{A_0}} \langle
N_1|O|N_0\rangle,\hspace{.7cm} B_2 = \frac{A_1}{A_0} \langle N_1|O|N_1\rangle.
\end{equation}
Fig.~\ref{fig_A4} shows a typical fit for ensemble
IV. The contributions to $C_{{\rm 3pt}}$ from excited states are large
for $t_{\rm f}=7a$, however, they steadily reduce, so that for $t_{\rm
  f}\ge 11a$ the results are consistent around $t\sim t_{\rm
  f}/2$. For $t_{\rm f}\ge 15a$ one can reasonably identify a plateau
over several timeslices. The fit reproduces the data well, with
reasonable values of $\chi^2/d.o.f. <2$, where correlations between
timeslices and the different correlators are taken into account. To
avoid any possible bias from an ill-determined covariance matrix, all
final results are taken from uncorrelated fits.  The systematic
uncertainty in $\langle x\rangle_{u-d}$ arising from the choice of fit
is estimated by varying the fitting range for both $C_{\rm
  2pt}$~($t_{\rm min}$ to $t_{\rm max}$) and $C_{\rm 3pt}$~($\delta t$
to $t_{\rm f}-\delta t$) where $\delta t, t_{\rm min}\ge 2a$ is allowed.
The number of $t_{\rm f}$s used in the fit was also varied.  Note that
the fitted value for $\langle x\rangle_{u-d}^{\overline{\rm
    MS}}(2\,\mathrm{GeV})$ indicated in Fig.~\ref{fig_A4}~(the green shaded region) only corresponds to a single
fit and the errors are purely statistical.

\begin{figure}[t]
\centerline{
\includegraphics[width=.48\textwidth,clip=]{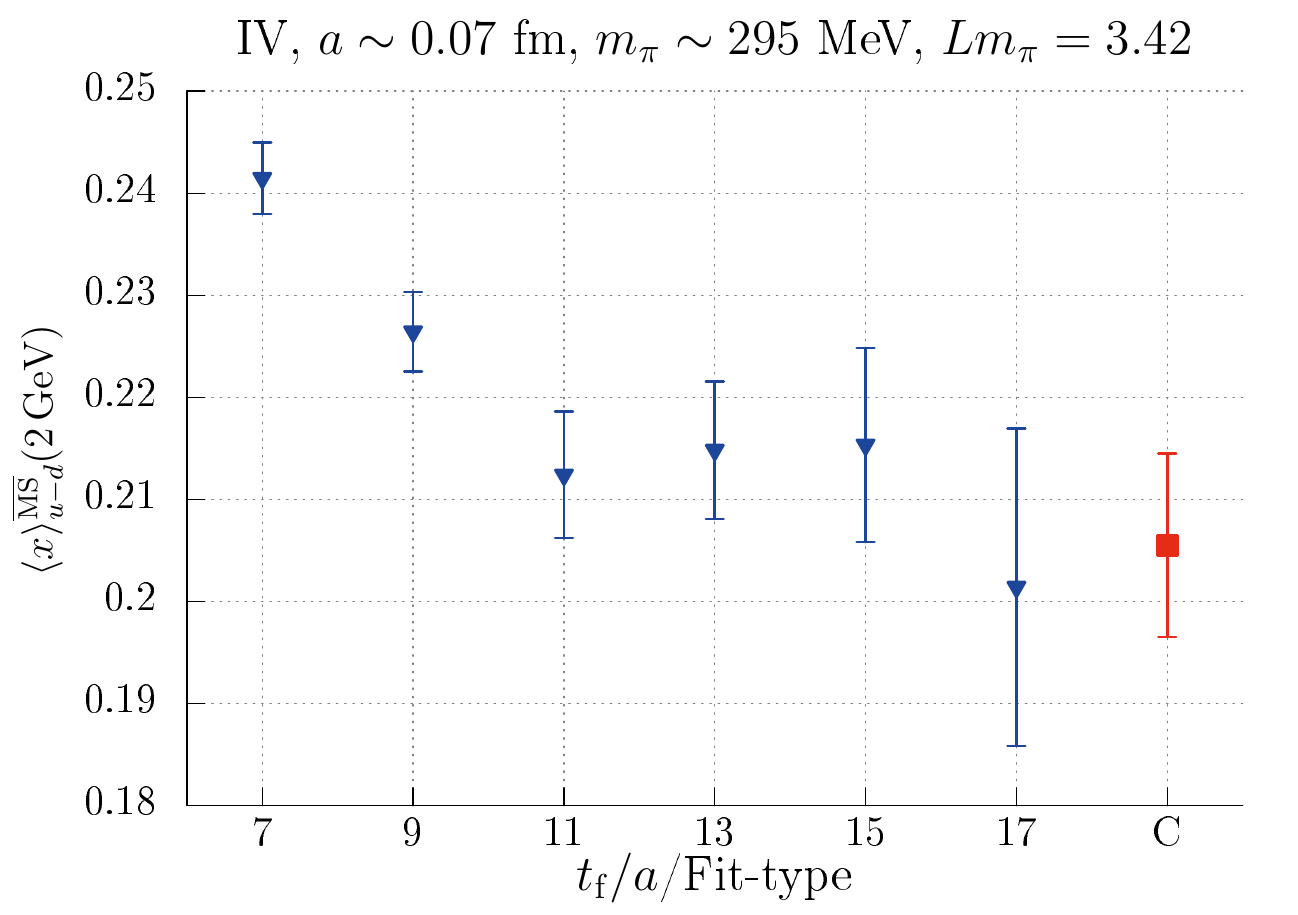}
}
\caption{Comparison of the values for $\langle
  x\rangle_{u-d}^{\overline{\rm MS}}(2\,\mathrm{GeV})$ extracted using
  constant fits to $C_{{\rm 3pt}}/C_{{\rm 2pt}}$ for different $t_{\rm f}$s with
  the result of a combined~(simultaneous) fit~($C$) to $C_{{\rm 3pt}}$ and
  $C_{{\rm 2pt}}$ for all $t_{\rm f}$, for ensemble IV~($m_\pi\sim 295$~MeV). }
\label{fig_A16}
\end{figure}

\begin{figure}[t]
\centerline{
\includegraphics[width=.48\textwidth,clip=]{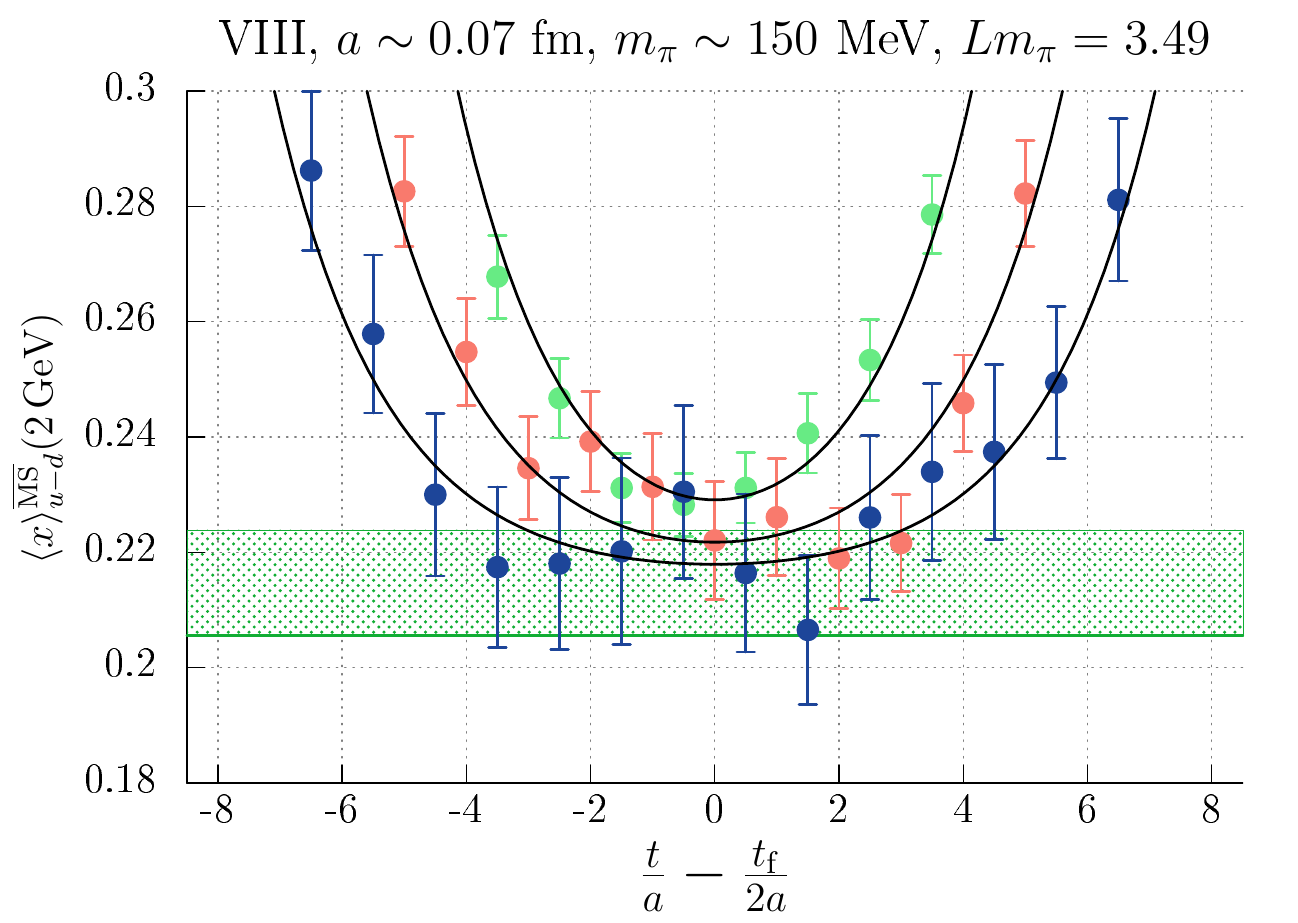}}
\caption{The same as Fig. \ref{fig_A4} for ensemble VIII~($m_\pi\sim 150$~MeV).
The same fitting ranges are employed.}
\label{fig_A8}
\end{figure}

\begin{figure}[t]
\centerline{
\includegraphics[width=.48\textwidth,clip=]{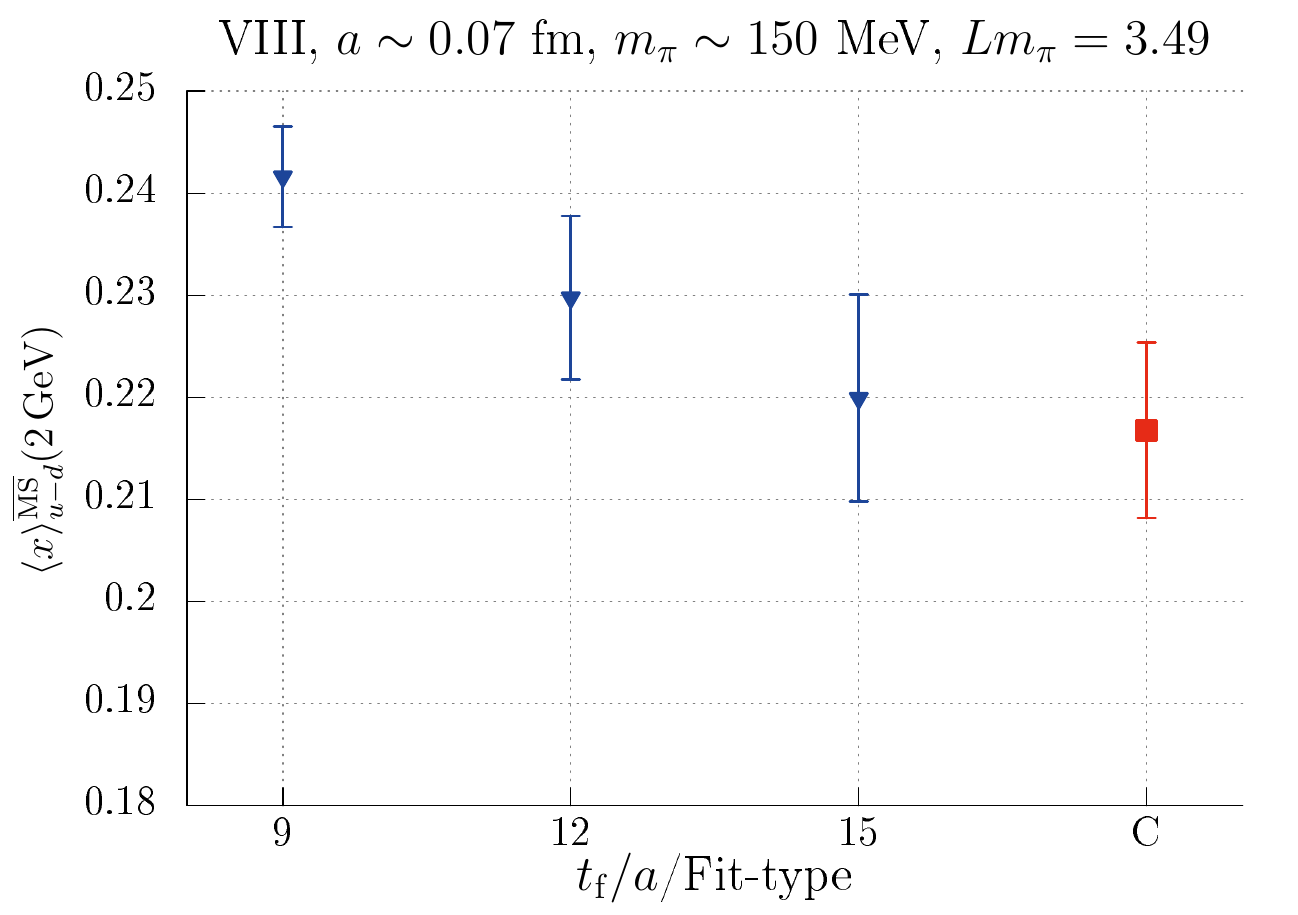}
}
\caption{The same as Fig. \ref{fig_A16} for ensemble VIII~($m_\pi\sim 150$~MeV).}
\label{fig_A19}
\end{figure}

The above approach, which we call ``combined'' fits, can be compared
to the traditional method of fitting the ratio $C_{{\rm
    3pt}}(t,t_{\mathrm{f}}) /C_{{\rm 2pt}}(t_{\mathrm{f}})$, for fixed
$t_{\mathrm{f}}$, to a constant $B_0$,
cf. Eq.~(\ref{ratiothreetwo}). In Fig.~\ref{fig_A16} one can see that
for $t_{\mathrm{f}}\ge 11a$ the results for the quark momentum
fraction are consistent with each other and are also consistent with
the results of the combined fits. Figs.~\ref{fig_A8} and~\ref{fig_A19}
show the corresponding results for ensemble VIII. In particular,
Fig.~\ref{fig_A19} suggests $t_{\mathrm{f}}=15a$ is sufficient for
suppressing excited state contributions and obtaining ground state
dominance for $t$ close to $t_{\rm f}/2$ at the present level
of statistical errors. However, this conclusion is only possible
through the use of optimized smearing and our extensive analysis.

\begin{figure}[t]
\centerline{
\includegraphics[width=.48\textwidth,clip=]{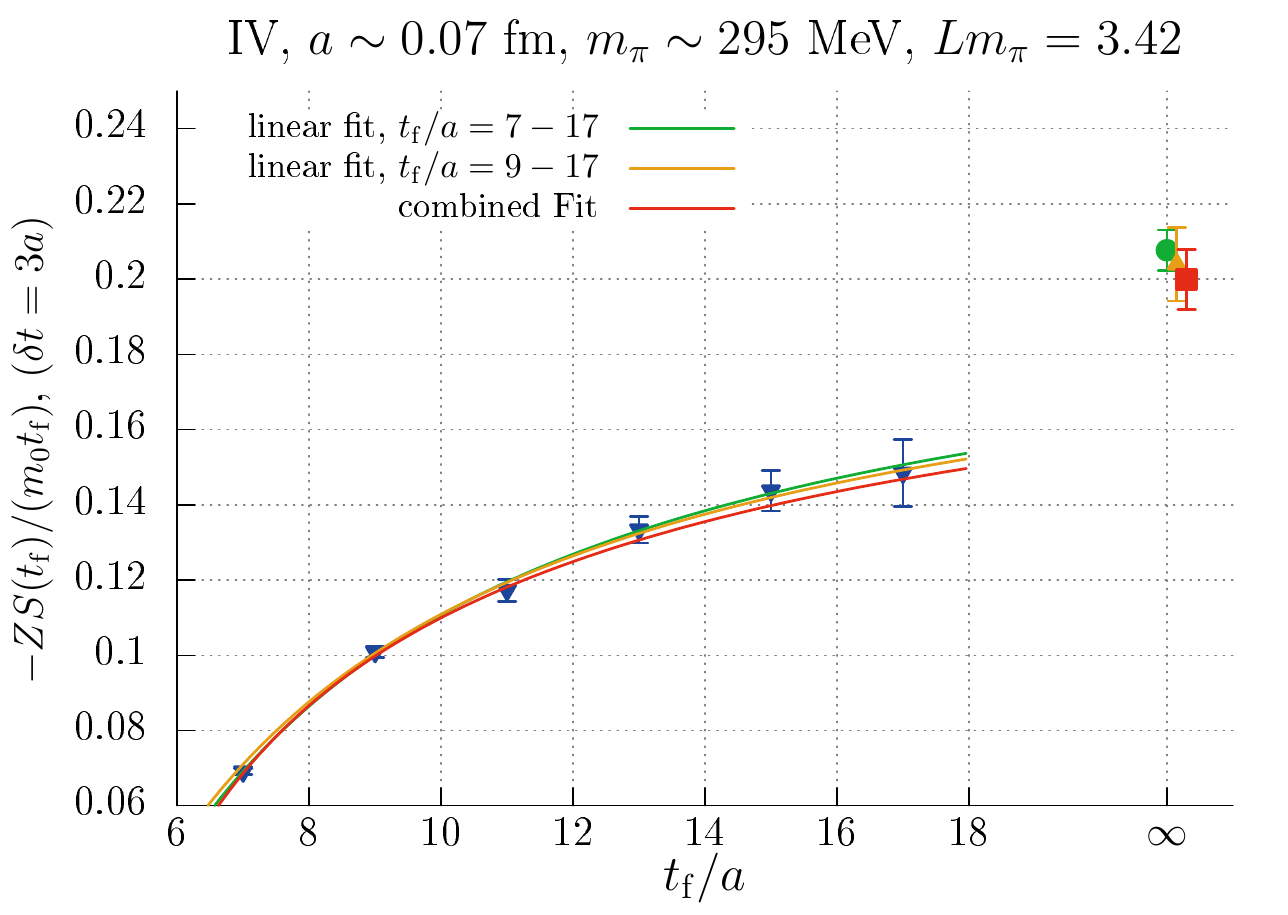}
}
\caption{$-ZS(t_{\rm f})/(m_0t_{\rm f})$~(blue triangles) for $\delta
  t=3a$ for ensemble IV~($m_\pi\sim 295$~MeV), which tends to $\langle
  x\rangle_{u-d}^{\overline{\rm MS}}(2\,\mathrm{GeV})$ as $t_{\rm
    f}\to\infty$. $Z$ corresponds to the renormalization factors in
  Eq.~(\ref{renorm_eq}). A comparison is made between the linear fits
  to $S(t_{\rm f})$ including different ranges of $t_{\rm f}$ values
  and $S(t_{\rm f})$ obtained using the parameters from a single
  combined fit~(shown in Fig.~\ref{fig_A4}) and Eq.~(\ref{sumeq}). The
  values of $\langle x\rangle_{u-d}^{\overline{\rm
      MS}}(2\,\mathrm{GeV})$ extracted from linear fits using $t_{\rm
    f}/a=7-17$~(green circle) and $t_{\rm f}/a=9-17$~(yellow triangle) and
  the combined fit~(red square) are shown on the right.}
\label{fig_A22}
\end{figure}

For completeness we also considered the summation
method~\cite{Maiani:1987by}, which has been been advertised in several
recent studies~\cite{Deka:2013zha,Doi:2009sq,Capitani:2010sg,Bulava:2010ej,Green:2012ud}.
This involves summing the ratio of the
three-point and two-point functions over a range of $t$ values:
\begin{equation}
S(t_{\rm f}) = \sum_{t=\delta t}^{t_{\rm f}-\delta t} C_{{\rm 3pt}}(t,t_{\rm f})/C_{{\rm 2pt}}(t_{\rm f}).\label{summ}
\end{equation}
Using Eqs.~(\ref{twopta}) and~(\ref{threepta}) one can show that
\begin{eqnarray}
S(t_{\rm f}) &=& \sum_{t=\delta t}^{t_{\rm f}-\delta t}\Bigl[ B_0+ B_1\Bigl( e^{-\Delta m(t_{\rm f}-t)} +
e^{-\Delta m\, t} \Bigr) 
\nonumber \\
&+&  B_2e^{-\Delta m\, t_{\rm f}} \Bigr] \Bigl[1+A_1e^{\Delta m\, t_{\rm f}}/A_0 \Bigr]^{-1}\label{sumeq}\\
& = & B_0t_{\rm f}+C+{\cal O}\Bigl( t_{\rm f}e^{-\Delta m\, t_{\rm f}}\Bigr),
\end{eqnarray}
where $C$ contains $t_{\rm f}$-independent terms. Thus, one can
extract $B_0$ by performing a linear fit to $S(t_{\rm f})$ as a
function of $t_{\rm f}$. A large number of $t_{\rm f}$ values are
required in order to obtain reliable results with this approach, where
the $e^{-\Delta m\, t_{\rm f}/2}$ corrections of the traditional ratio
method were traded in against $1/t_{\rm f}$ corrections for the slope.  We
compare the summation method and the combined fit approach using
ensemble IV in Fig.~\ref{fig_A22}. In this case we chose $\delta t=3a$~(cf.
Eq.~(\ref{summ})) to minimize the contributions from excited
states. Similarly, one can omit the results with the smallest $t_{\rm
  f}$ in the fit. However, we found consistent results for $B_0$, with
and without the $t_{\rm f}=7a$ points. Agreement is also seen with
$S(t_{\rm f})$ obtained using Eq.~(\ref{sumeq}) and the parameters
determined from the combined fit.  No advantage was found in using the
summation method, in particular, given the need for many $t_{\rm f}$
values in order to confirm the linear behaviour of $S(t_{\rm f})$.

\begin{figure}[t]
\centerline{
\includegraphics[width=.48\textwidth,clip=]{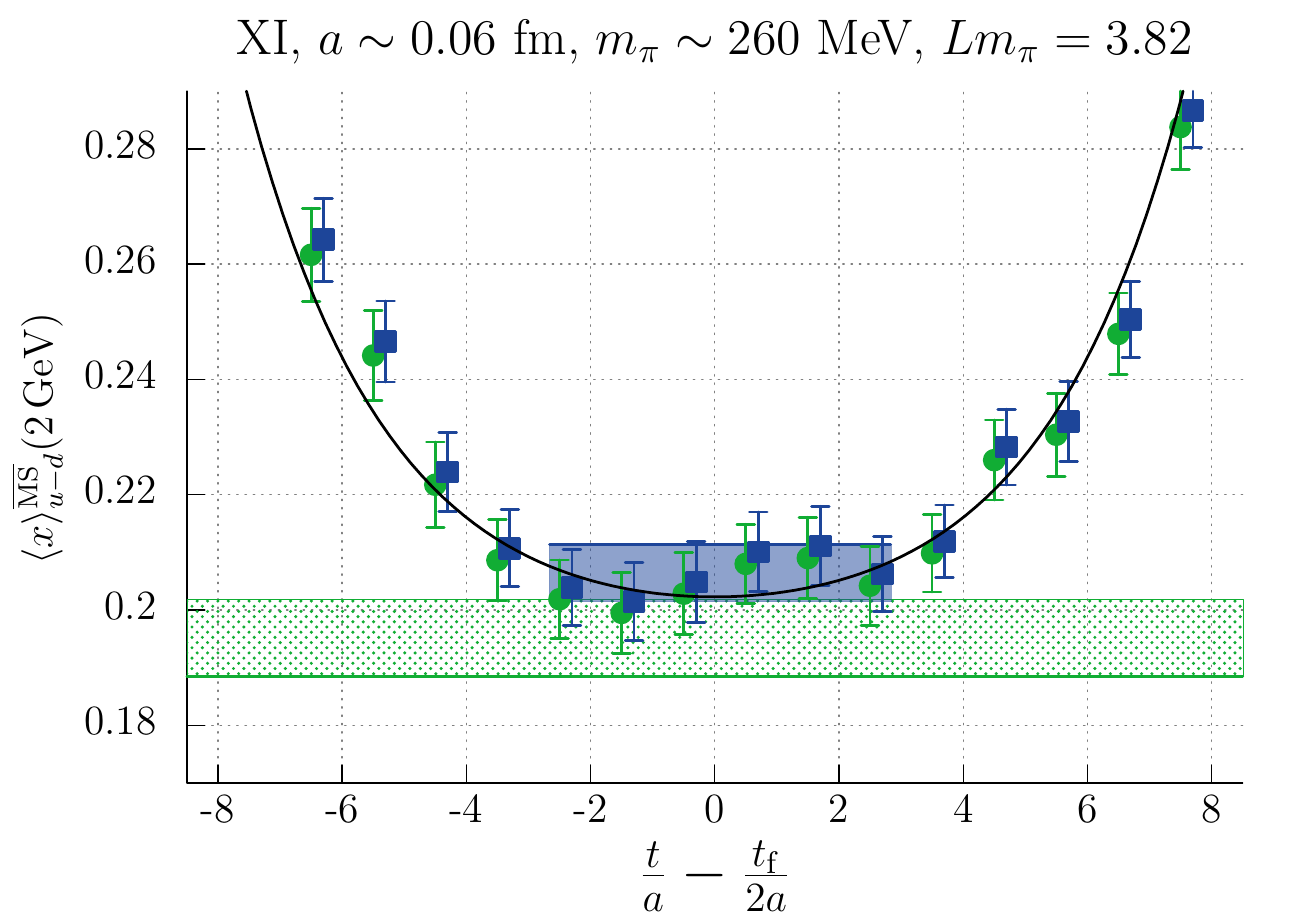}}
\caption{The same as Fig. \ref{fig_A4}~(green circles and shaded
  region) for ensemble XI~(similar $m_\pi$ but $a\sim 0.06$~fm
  rather than $a \sim 0.07$~fm). The combined fit is performed with the ranges
  $t_{\rm min}-t_{\rm max}=2a-26a$ for $C_{\rm 2pt}$ and $\delta t =2a$ for
  $C_{\rm 3pt}$. In addition,  the ratio $C_{{\rm 3pt}}/C_{{\rm 2pt}}$,
  including the appropriate factors to give $\langle
  x\rangle_{u-d}^{\overline{\rm MS}}(2\,\mathrm{GeV})$~(blue squares)
  and the result of a constant fit to this ratio in the range
  $t/a=6-11$~(blue shaded region) are shown. }
\label{fig_A11}
\end{figure}

For the other ensembles the three-point functions were computed with
only one value $t_{\rm f}=15a$, rescaled for different lattice
spacings at $\beta=5.20$~($t_{\rm f}=13a$) and $\beta=5.40$~($t_{\rm
  f}=17a$). 
This is justified by the observation that 
\begin{equation}
\frac{|B_2|e^{-\Delta m\cdot 1\mathrm{fm}}}{|B_0|} < 2\cdot 10^{-3}
\end{equation}
on ensembles IV and VIII.
We performed combined fits of the $C_{{\rm 3pt}}$ and
$C_{{\rm 2pt}}$ using Eqs.~(\ref{twopta}) and~(\ref{threepta}) setting
$B_2=0$.  A typical example is shown in Fig.~\ref{fig_A11}.  Also
included in the figure for comparison is the ratio $C_{{\rm 3pt}}/C_{{\rm 2pt}}$ and the result for a constant fit to this
ratio. Reasonable agreement is found between both fitting methods.

As a consistency check we also utilized the value of $B_2$ 
determined in the fits to ensemble IV:
\begin{equation}
B_2 = \frac{A_1}{A_0} \langle N_1|O|N_1\rangle = -m_1\frac{A_1}{A_0} \tilde B_2
\end{equation}
where $\tilde B_2$ corresponds to $\langle x\rangle_{u-d}$ for the
first excited state.  Assuming a weak dependence of this on
$m_\pi$~(as is the case for the ground state), the $B_2$ term in
Eq.~(\ref{threepta}) can be replaced by the r.h.s. above and $\tilde
B_2$ can be fixed to the value obtained from ensemble IV.  However,
$a\Delta m$ is typically around $0.5$, which means this term is very
small at $t_{\rm f}\ge 13a$, and no noticeable changes occurred in the
fit results.

\section{Extrapolation of the iso-vector generalized form factor $A_{20}$}
\label{finitemom}

As a further check we extracted the~(iso-vector)
generalized form factor, $A_{20}(q^2)$, which in the forward limit is
equal to $\langle x\rangle_{u-d}$. The generalized form factor appears in
the Lorentz decomposition of the matrix element~(in Euclidean
  notation)
\begin{eqnarray}
&&\langle N(p_{\rm f})| \bar{q} \gamma_{\{\mu}\!\!\stackrel{\leftrightarrow}{D}\!\!{}_{\nu\}}q|N(p)\rangle
 \nonumber\\ && = S(\mu,\nu)\left[\bar{u}(p_{\rm f})\left(i\gamma_\mu \overline{p}_\nu A_{20}(Q^2)
+i\sigma_{\mu\alpha}\frac{q_\alpha \overline{p}_\nu}{2m_N}B_{20}(Q^2)
\right.\right.
\nonumber\\ && \left.\left. \hspace{3.5cm}
+ \frac{q_\mu q_\nu}{m_N}C_{20}(Q^2)\right)u(p)\right],\label{generalff}
\end{eqnarray}
where $u(p)$ and $\bar{u}(p_{\rm f})$ are fermion and anti-fermion
spinors, respectively, and $\overline{p}=\frac{1}{2}(p_{\rm f}+p)$,
$q=(p_{\rm f}-p)$ and $Q^2=-q^2$.  The indices $\mu$, $\nu$ are
symmetrized on both sides of the equation (indicated by the curly
brackets on the l.h.s.\ and the symmetrization function $S(\mu,\nu)$ on
the r.h.s.) and the traces are subtracted. One constructs
combinations of $O_{\mu\nu}=
\bar{q}\gamma_{\{\mu}\!\!\stackrel{\leftrightarrow}{D}\!\!{}_{\nu\}}q$ which
correspond to irreducible representations of the hypercubic group to
avoid mixing under renormalization.

The corresponding matrix elements can be extracted using the following
ratio of finite momentum three-point and two-point functions in the
region of ground state dominance:
\begin{eqnarray}
\frac{C_{\rm 3pt}(t,t_{\rm f};\vec{p},\vec{p}_{\rm f})}{C_{\rm 2pt}(t_{\rm f};\vec{p}_{\rm f})}
\left[\frac{C_{\rm 2pt}(t;\vec{p}_{\rm f}) C_{\rm 2pt}(t_{\rm f};\vec{p}_{\rm f}) C_{\rm 2pt}(t_{\rm f}-t;\vec{p})}{C_{\rm 2pt}(t;\vec{p}) C_{\rm 2pt}(t_{\rm f};\vec{p}) C_{\rm 2pt}(t_{\rm f}-t;\vec{p}_{\rm f})}\right]^{\frac{1}{2}},\nonumber\\
\label{ratio_finitemom}
\end{eqnarray}
where
\begin{eqnarray}
C_{{\rm 2pt}}(t_{\mathrm{f}};\vec{p}) &= & \sum_{\vec{x}}e^{-i\vec{p} \cdot\vec{x}}\langle {\cal N}(\vec{x},t_{\mathrm{f}})\overline{ {\cal N}}(\vec{0},0)\rangle \\
C_{{\rm 3pt}}(t_{\mathrm{f}},t;\vec{p}_{\rm f},\vec{p}) & = &\sum_{\vec{x},\vec{y}} e^{-i\vec{p}\cdot\vec{x}+i(\vec{p}_{\rm f}-\vec{p})\cdot \vec{y}}\nonumber\\
&&\times\langle {\cal N}(\vec{x},t_{\mathrm{f}}) O(\vec{y},t)\overline{ {\cal N}}(\vec{0},0)\rangle.
\end{eqnarray}
Further details and results on generalized form factors will be
provided in a forthcoming publication, including an analysis of the
excited state contributions.

\begin{figure}[t]
\centerline{
  \includegraphics[width=.48\textwidth,clip=]{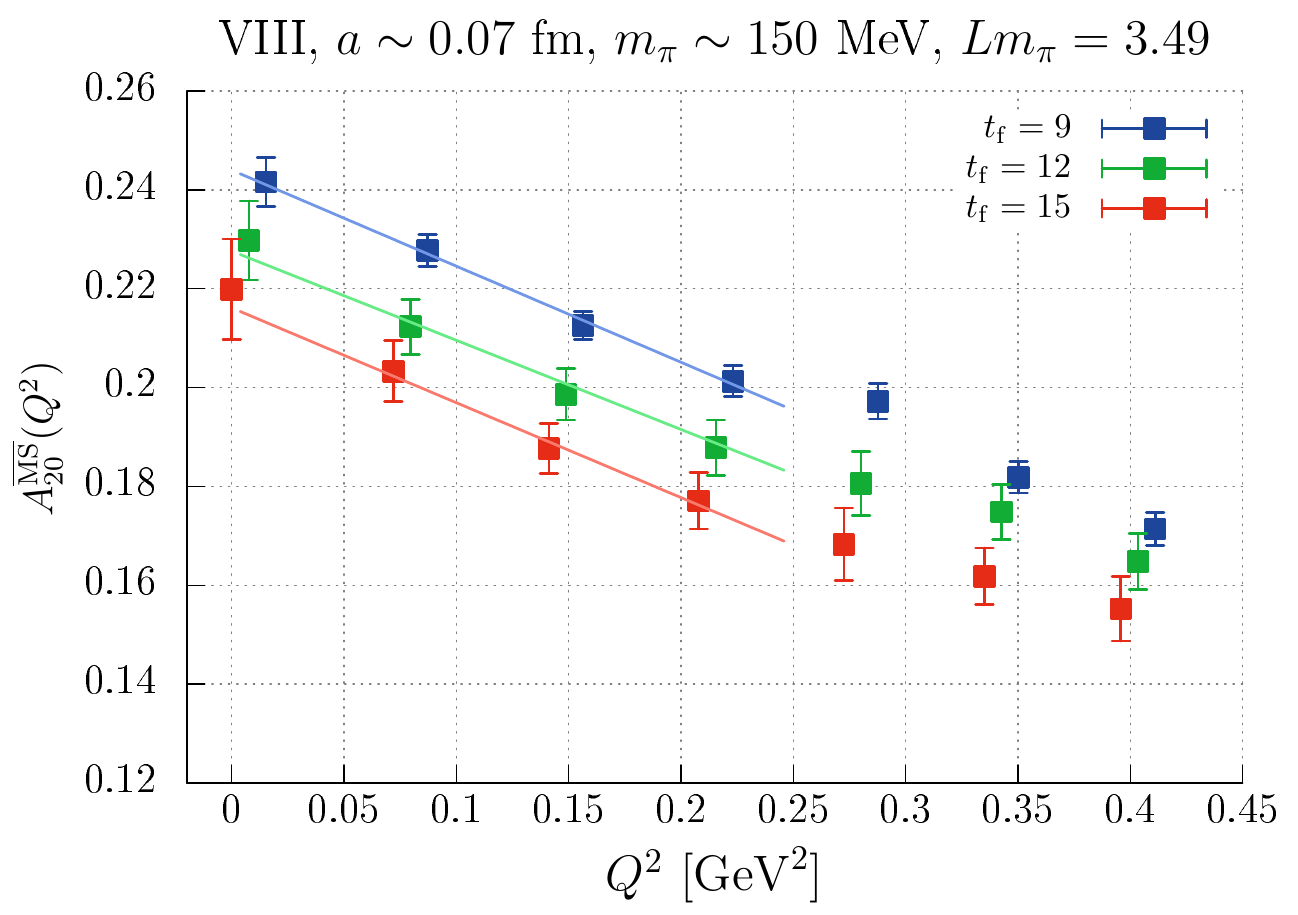}
}
\caption{The iso-vector generalized form factor
  $A^{\overline{\mathrm{MS}}}_{20}~(Q^2)$ at $\mu=2$~GeV for ensemble
  VIII~($m_\pi\sim 150$~MeV) extracted using
  Eq.~(\ref{ratio_finitemom}) with different $t_{\rm f}$ values. The
  solid lines indicate linear extrapolations, for fixed $t_{\rm f}$,
  to the forward limit~(not including the $q^2=0$ points). The data
  for different $t_{\rm f}$s are shifted horizontally for clarity. }
\label{fig_a20_light}
\end{figure}

Figure~\ref{fig_a20_light} shows
$A^{\overline{\mathrm{MS}}}_{20}~(Q^2)$ at $\mu=2$~GeV for ensemble
VIII for the three values of $t_{\rm f}$. In general, excited state
contributions depend on the momentum of the correlation functions. We
find the dependence on $t_{\rm f}$ to be similar to that in the
forward limit for the lowest three values of $Q^2$. Performing a
linear extrapolation in $Q^2$, as suggested, e.g., by leading order
chiral perturbation theory, Fig.~\ref{fig_a20_light} shows we obtain
consistency with our analysis at $Q^2=0$. In addition, we found
agreement between results obtained using different hypercubic
irreducible representations, which are sensitive to different
discretization effects, albeit within large statistical uncertainties.

\section{Results}
\label{sys}

\begin{figure}[t]
\centerline{
\includegraphics[width=.48\textwidth,clip=]{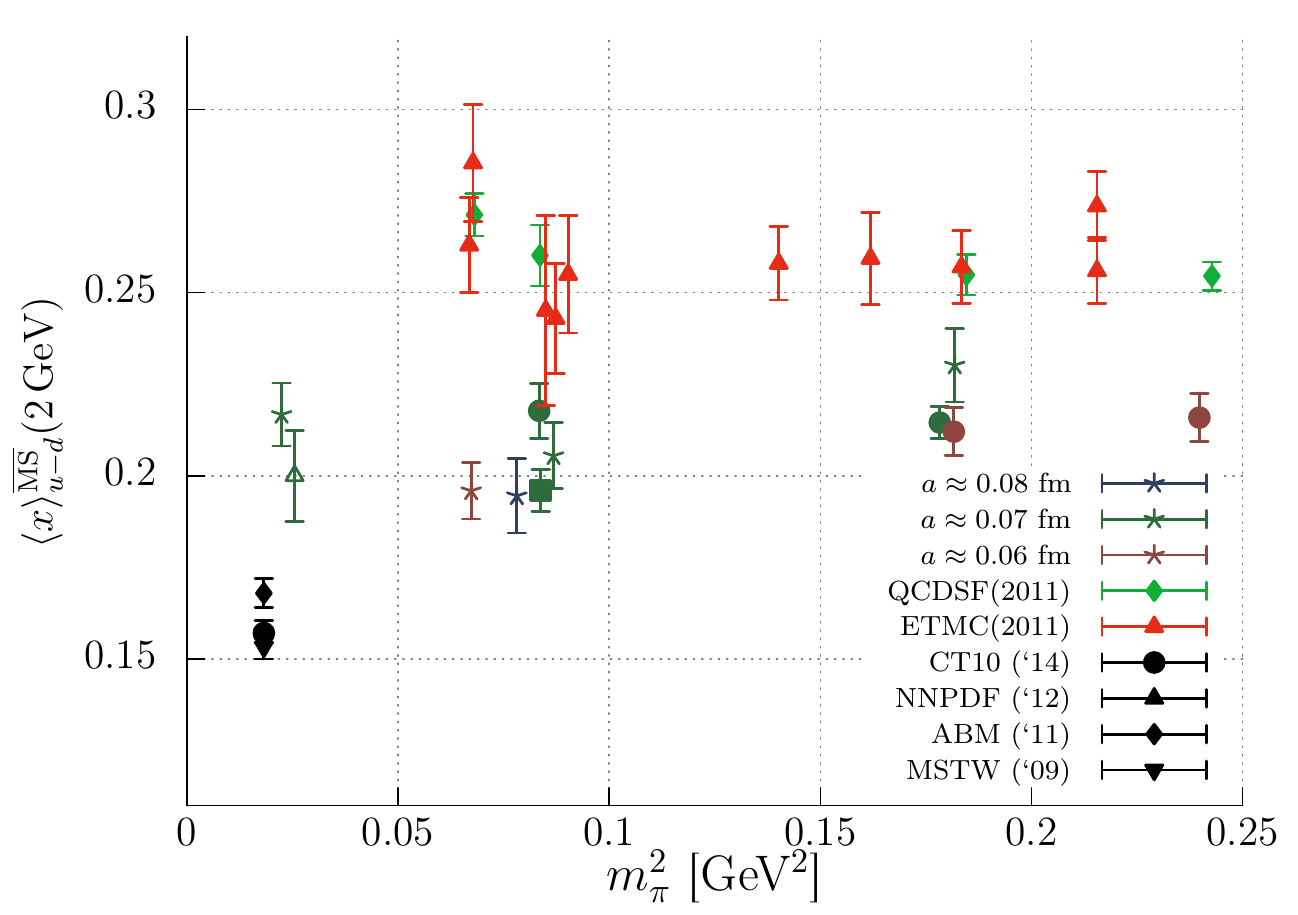}
}
\caption{Our results for $\langle x\rangle_{u-d}^{\overline{\rm MS}}$ for
  the three lattice spacings, $a=0.08$~fm~(blue), $a=0.07$~fm~(green)
  and $a=0.06$~fm~(red), and in terms of the volume,
  $L m_\pi<3.4$~(triangles), $3.4\le L m_\pi<4.0$~(crosses), $4.0 \le
  L m_\pi<6.0$~(circles) and $L m_\pi>6.0$~(squares).  For
  comparison, earlier $N_f=2$ results from
  QCDSF~\cite{Pleiter:2011gw,Sternbeck:2012rw} and
  ETMC~\cite{Alexandrou:2011nr} are included, as well as the values
  obtained from phenomenological fits to experimental data by
  CT10~\cite{Gao:2013xoa}, ABM~\cite{Alekhin:2013nda},
  NNPDF~\cite{Ball:2012cx} and MSTW~\cite{Martin:2009bu}. }
\label{fig_A23}
\end{figure}

\begin{figure}[h]
\centerline{
\includegraphics[width=.48\textwidth,clip=]{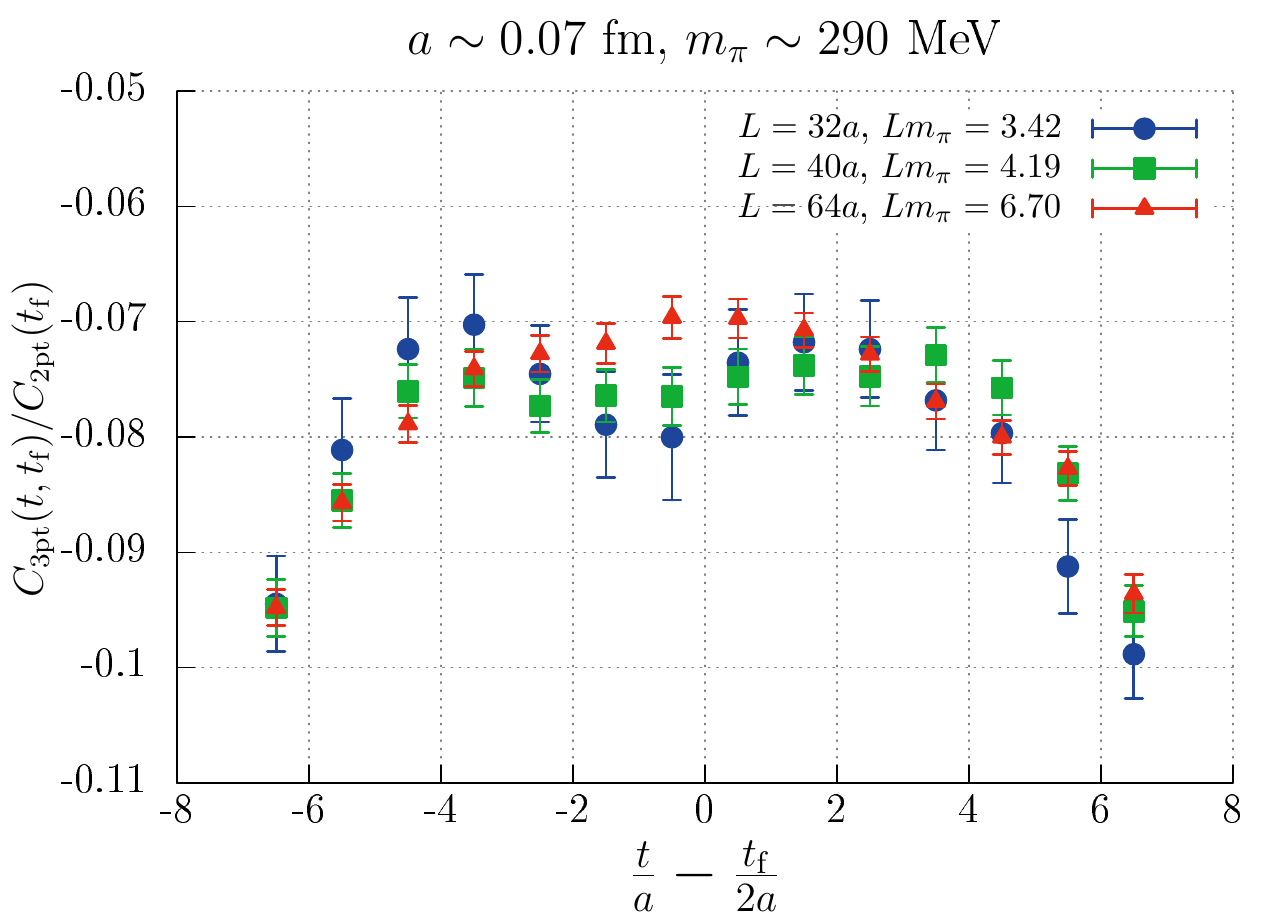}
}
\caption{The ratio of the three-point to two-point functions for
  ensembles IV, V and VI, which have the same $\beta$ and $\kappa$
  values~ but have different spatial extents $L$. $m_\pi\sim
  290$~MeV for $L=64a$.}
\label{fig_A17}
\end{figure}

\begin{figure}[h]
\centerline{
\includegraphics[width=.48\textwidth,clip=]{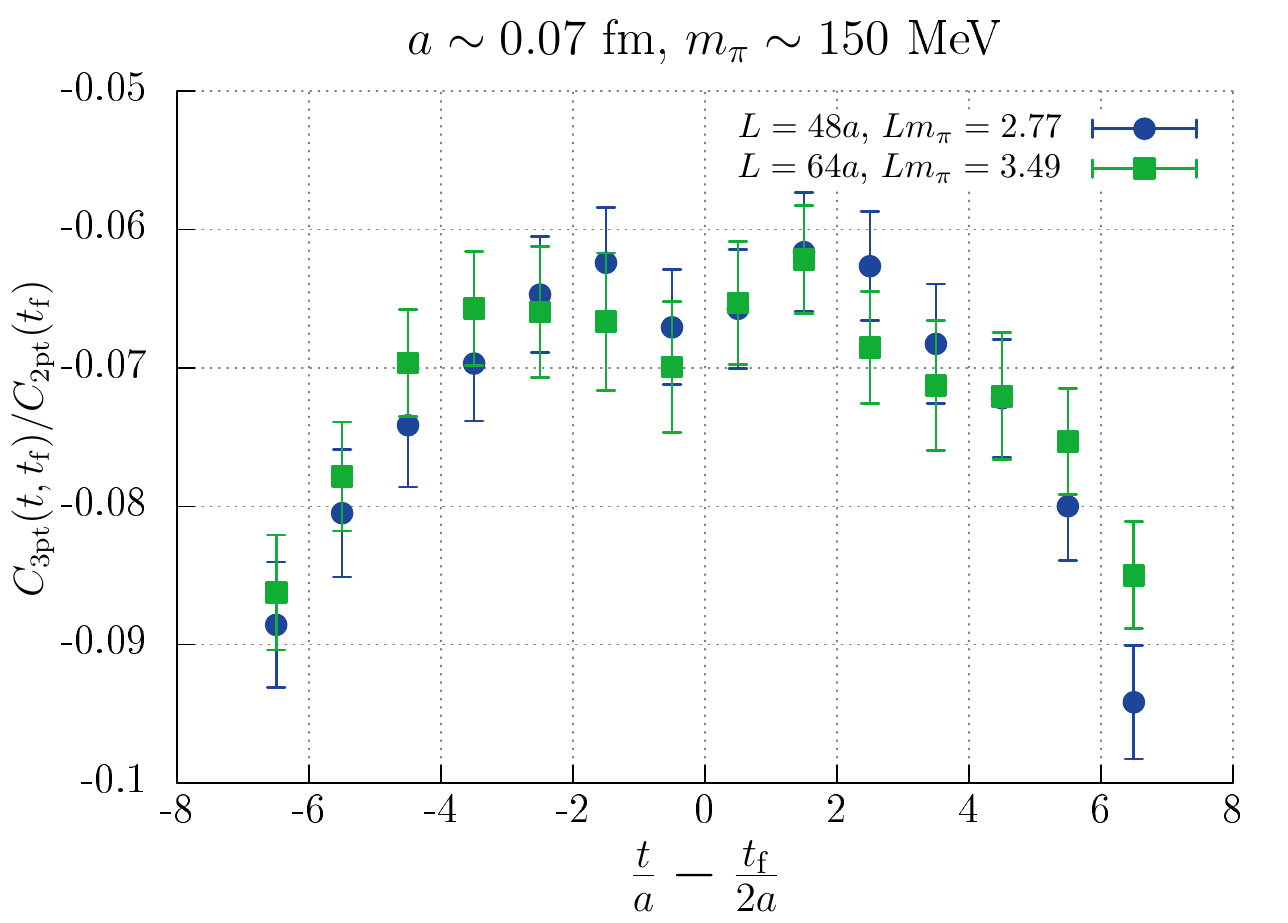}
}
\caption{The same as in Fig.~\ref{fig_A17} for ensembles VII and VIII.
$m_\pi\sim
 150$~MeV for $L=64a$.}
\label{fig_A20}
\end{figure}

Our results for $\langle x\rangle_{u-d}^{\overline{\rm MS}}$ at the
scale $\mu=2$~GeV for all ensembles are given in Fig.~\ref{fig_A23} as
a function of $m_\pi^2$. Strikingly, we see that there is no
significant dependence on the quark mass, neither for the
range $m_\pi\sim 490-289$~MeV and  $4.0\le L m_\pi\le 6.0$~(the circles
in the figure) nor for $m_\pi\sim 295-150$~MeV and $3.4\le L m_\pi\le
4.0$~(crosses). At the three values of $m_\pi$ in the range
$422-150$~MeV, where we have more than one volume, we also find no
significant finite volume effects.  To emphasize this point we show
the ratio of the three-point to the two-point
functions, Eq.~(\ref{ratiothreetwo}), in Figs.~\ref{fig_A17}
and~\ref{fig_A20} for $m_\pi\sim 295$~MeV and $m_\pi\sim 150$~MeV,
respectively. For the larger pion mass, the three volumes correspond
to $L m_\pi \sim 3.42-6.70$ while at the near physical pion mass $L m_\pi$ is
only varied in the range $2.77-3.49$. However, at fixed $L m_\pi$, if
the finite volume effects arise from pion exchange, the relative
finite volume correction is proportional to $m_\pi^2$. Thus, we do not
expect the emergence of significant finite volume effects for larger
volumes for $m_\pi\sim 150$~MeV. In terms of lattice spacing effects,
our analysis is not conclusive. Although no significant effects are seen,
we have leading $O(a)$ discretization errors and the lattice spacing is
only varied in a very limited range $a\sim 0.08-0.06$~fm.

In Ref.~\cite{Bali:2012av} we compared our results for ensemble VII,
$m_\pi\sim 160$~MeV~(with lower statistics) with earlier, much larger,
$N_f=2$ results, also shown in Fig.~\ref{fig_A23}, which used similar
analyses and~(non-perturbative)
renormalization but different smearing.  At that time, this suggested
a strong dependence on $m_\pi$ as one approaches the chiral limit.
However, from our present analysis including larger pion
masses, optimized smearing and excited state fits throughout, we conclude the
observed difference is probably due to excited state
contamination~(cf. Fig.~\ref{fig_A21}). Nonetheless there remains a
$\sim 25\%$ discrepancy with the values obtained from phenomenological
fits to the experimental
data\footnote{Note that these fits are only performed
at NNLO while we converted the lattice results from the
RI'MOM to the $\overline{\rm MS}$ scheme at three-loop order. We also
evolved the scale to this order. However, differences
between running the scale to 2~GeV at
two-, three- or four-loops~\cite{Velizhanin:2011es} are only on the few
per mille level.}~\cite{Gao:2013xoa,Alekhin:2013nda,Ball:2012cx,Martin:2009bu}.

\begin{figure}[t]
\centerline{
\includegraphics[width=.48\textwidth,clip=]{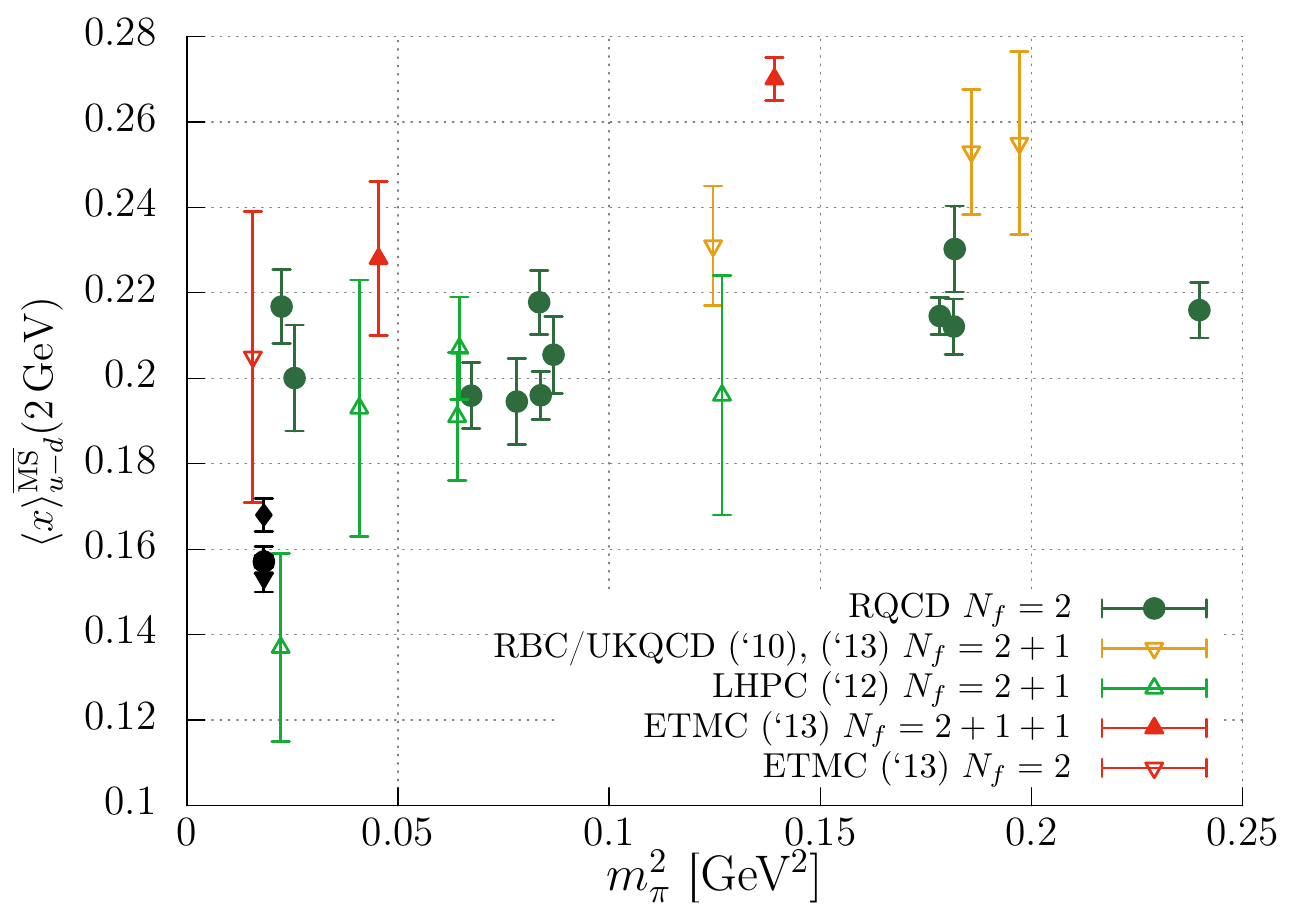}
}
\caption{Comparison with recent $N_f=2$, $2+1$ and $2+1+1$ simulations
  from LHPC~\cite{Green:2012ud}, RBC/UKQCD~\cite{Aoki:2010xg} and
  ETMC~\cite{Alexandrou:2013jsa,Alexandrou:2013joa}. The results from 
phenomenological fits~(black points) are the same as in Fig.~\ref{fig_A23}. }
\label{fig_C1}
\end{figure}

In Fig.~\ref{fig_C1} a comparison is made with recent determinations
employing $N_f=2+1$ dynamical fermions~(LHPC~\cite{Green:2012ud} using
tree-level improved clover fermions with 2-HEX link smearing and
RBC/UKQCD~\cite{Aoki:2010xg} with domain wall fermions) and with
$N_f=2+1+1$ and $N_f=2$
simulations~(ETMC~\cite{Alexandrou:2013jsa,Alexandrou:2013joa} using
twisted mass fermions). All collaborations use non-perturbative
renormalization and the unimproved lattice operator. Overall, within
the larger errors of these collaborations, consistency can be seen
with our results~(one high statistics ETMC point being the only
exception).  Higher precision is needed to resolve any effects of
including strange quarks in the sea or to uncover discretization
effects.  For instance, LHPC~\cite{Green:2012ud} reports agreement
with the phenomenological value at almost physical pion masses using,
predominantly, coarse $a\sim 0.12$~fm ensembles, however, 
within quite large errors, see Fig.~\ref{fig_C1}.

\section{Conclusions}
\label{conc}

In this article we presented high statistics results for the
iso-vector quark momentum fraction, $\langle x\rangle_{u-d}$, with
pion masses down to $150$~MeV and with volumes up to $L m_\pi>6$. This
quantity is sensitive to excited state contributions and through the
use of optimized smearing, multiple sink time positions and excited
state fits we were able to extract ground state signals
unambiguously. No significant dependence was observed on the quark
mass within the range, $150\,\mathrm{MeV}<m_\pi<490\,\mathrm{MeV}$,
nor on the lattice volume, even close to the physical point.  The
consistency found with other recent determinations suggests that
strange sea quarks do not play an important role for this valence
quantity. The remaining discrepancy between our result $\langle
x\rangle_{u-d}^{\overline{\mathrm{MS}}}(2\,\mathrm{GeV})=0.217(9)$ obtained at
$m_{\pi}\sim 150$~MeV and $a\sim 0.071$~fm and phenomenological values
$\sim 0.15$--0.17 may be due to discretization effects. At present, we do
not have these under control with $O(a)$ leading lattice
spacing effects and only a small variation in $a$. We remark that all
lattice studies have leading order $a$ discretization effects for this
quantity and lattice spacings $a\gtrsim 0.06$~fm in common.  We are
also in the process of revisiting the determination of the
non-perturbative renormalization factor since this has been computed
using similar methods in all recent lattice investigations.  In the
future we plan to realize $a<0.06$~fm, simulating $N_f=2+1$ sea quarks
with open boundaries~\cite{Luscher:2012av}.

\section*{Acknowledgments}

The ensembles were generated primarily on the QPACE
computer~\cite{Baier:2009yq,Nakamura:2011cd}, which was built as part
of the Deutsche Forschungs Gemeinschaft SFB/TRR 55 project.  The
analyses were performed on the iDataCool cluster in Regensburg and the
superMUC system at the Leibniz Supercomputing Center in
Munich. Additional support was provided by the EU: ITN STRONGnet and
FP7-PEOPLE-2009-IRG, No.256594. The BQCD~\cite{Nakamura:2010qh} and
CHROMA~\cite{Edwards:2004sx} software packages were used extensively
along with the locally deflated domain decomposition solver
implementation of openQCD~\cite{luscher3}.

\bibliography{xumd}

\end{document}